\documentclass[aps,twocolumn,showpacs,superscriptaddress]{revtex4-2}
\usepackage{amsmath,amssymb}
\usepackage{mathrsfs}
\usepackage{graphics,graphicx}
\usepackage{dcolumn,bm}
\usepackage{breqn}
\usepackage{psfrag}
\usepackage{enumitem}
\usepackage{color}
\usepackage{enumitem}
\usepackage [english]{babel}
\usepackage [autostyle, english = american]{csquotes}
\usepackage{mathtools}
\DeclarePairedDelimiter\bra{\langle}{\rvert}
\DeclarePairedDelimiter\ket{\lvert}{\rangle}
\DeclarePairedDelimiterX\braket[2]{\langle}{\rangle}{#1 \delimsize\vert #2}

\MakeOuterQuote{"}
\usepackage{blkarray}
\newcommand{\cu}
{\affiliation{Department of Physics, University of Calcutta, 92 Acharya Prafulla Chandra Road, Kolkata 700009, India.}
\affiliation{Department of Physics, SRM University-AP, Amaravati, Andhra Pradesh - 522240, India.}}

\begin{document}
\title{ Emergent Quantum Walk Dynamics from Classical Interacting Particles }

\author{Surajit Saha }%
\cu

\normalsize  

\begin{abstract}

The dynamics of a discrete-time quantum walk (DTQW) can be realized within a
purely classical interacting particle system composed of some boxes and a
large but finite number of balls, and can, in principle, be implemented in a
tabletop experimental setting. The distribution of the balls evolves under stochastic, occupation-dependent update rules at each lattice site, producing quantum-walk dynamics without invoking a wavefunction. The update parameters are fixed by the parameters of coin and shift operations of the DTQW. This framework naturally yields a generalized active spin model and provides a minimal lattice-based microscopic understanding of the emergence of quantum-like dynamics in active matter systems. This interdisciplinary approach connects the classical models to the broad range of applications where DTQWs are successfully employed.

\end{abstract}

\pacs{}

\maketitle
Discrete-time quantum walks (DTQW) on graphs serve as a well-established framework for developing quantum algorithms. Though conceptually straightforward, they possess remarkable computational power and have been shown to be universal for quantum computation \cite{love,qwu}. Owing to their non-classical spreading dynamics, localization property, and potential algorithmic advantages, DTQW have been extensively explored both theoretically and experimentally \cite{Portugal2013, Kadian2021,Yuan2023}.\\

Modeling quantum phenomena like DTQWs on classical platforms is inherently challenging. This difficulty stems from the fact that DTQWs rely on quantum superposition, where evolving states interfere—constructively or destructively—based on their relative phases. As a result, they produce probability distributions that differ from those of classical random walks.

A line of research on this topic has approached this challenge by constructing classical random walks (CRWs) \cite{tomoki,mmro,yama} whose space and time-dependent transition probabilities in each step are extracted from the amplitudes of the DTQW. This approach yields a non-homogeneous CRW whose probability distribution replicates that of the corresponding DTQW. Although effective in reproducing certain statistics, these methods necessarily break the spatial and temporal structure of the DTQW and demand access to quantum amplitudes at each step, an approach analogous to Bohmian mechanics\cite{durr}, where the quantum wave function guides particle trajectories.

In contrast, alternative formulations of quantum mechanics—akin to the hydrodynamic approaches of Holland and Poirier \cite{holl,poir,parl,shiff,sebens,wiseman} —eschew direct reliance on the wave function, instead constructing real-valued fields or potentials that reproduce quantum evolution without amplitude-based guidance. Inspired by this distinction, we propose an interacting particles framework for modeling DTQWs that never invokes complex amplitudes. All update rules are real‑valued  processes whose parameters depend only on the parameters of Quantum walker’s coin and shift operations. Crucially, our model mirrors the structural properties of the original quantum walk system. When the Quantum walk has uniform coin and shift operations, the classical interacting particle dynamics remains homogeneous in space and time, employing identical update rules at each node and time step. If Quantum walker's coin or shift operations vary across space or time, the corresponding interacting particle model adapts its parameters in the same pattern, preserving the structural variations of Quantum walk. The Classical interacting particle model yields two main advantages. First, preserving the structural variations of the Quantum walk simplifies the analysis, while allowing our method to be naturally extended to higher-dimensional graphs. Second, by avoiding amplitude-based guidance, the classical model brings into focus a fundamental probabilistic mechanism underlying quantum walk spreading and interference, insights that may carry foundational significance.



To define the one-dimensional discrete-time quantum
walk we require the coin Hilbert space $\mathcal{H}_c$
and the position Hilbert space $\mathcal{H}_p$. The $\mathcal  H_{c}$ is spanned by the internal 
basis states of the particle, $|0\rangle$ and $|1\rangle$, and  the 
$\mathcal  H_{p}$ is spanned  by the position basis states $|n\rangle$, 
$n \in \mathbb{Z}$. The total system is then in the space $\mathcal H
= \mathcal  H_{c} \otimes \mathcal  H_{p}$. To implement  a simple
version of the quantum walk,  known as the Hadamard walk,
 one applies the Hadamard operator


\begin{equation}
  H = \frac{1}{\sqrt{2}}
      \begin{pmatrix}
        1 & 1 \\[4pt]
        1 & -1
      \end{pmatrix}
  \label{eq:hadamard_operator}
\end{equation}

on $\mathcal{H}_c$, which transforms a particle at the origin in one of the basis states into an equal superposition of $\ket{0}$ and $\ket{1}$.  
Next, we perform the conditional shift
\begin{equation}
  S = 
    \bigl(\ket{0}\bra{0}\otimes \sum_{n=-\infty}^{\infty}\ket{n+1}\bra{n}\bigr)
  + \bigl(\ket{1}\bra{1}\otimes \sum_{n=-\infty}^{\infty}\ket{n-1}\bra{n}\bigr).
  \label{eq:shift_operator}
\end{equation}
A single step of the Hadamard quantum walk is thus
\begin{equation}
  W \;=\; S \bigl(H \otimes \mathbb{I}_{p}\bigr),
  \label{eq:walk_operator}
\end{equation}
Here $\mathbb{I}_{p}$ is the identity operator on $\mathcal{H}_c$. 
The process of $W$ is iterated without resorting to the intermediate measurements to realize a large number of steps of the quantum walk.\\
It has been demonstrated in \cite{cmc1} that a three-parameter $SU(2)$ quantum coin
operation,
\begin{equation}
  C(\xi,\theta,\zeta) 
  = \begin{pmatrix}
      e^{i\xi}\cos\theta & e^{i\zeta}\sin\theta \\[6pt]
      -e^{-i\zeta}\sin\theta & e^{-i\xi}\cos\theta
    \end{pmatrix},
  \label{general_coin}
\end{equation}
is sufficient to describe the most general form of the discrete-time quantum walk,
where $\xi,\theta,\zeta\in\mathbb{R}$ and $\theta\in[0,\pi/2]$.  
By choosing different $(\xi,\theta,\zeta)$, one obtains a variety of discrete-time walks with distinct dynamical features. For example one can write $H= C(0,\frac{\pi}{4},0)$.

 
 Consider a general single-qubit quantum state given by
 \begin{equation}
\ket{\psi}=\psi_0 \ket{0} +\psi_1 \ket{1}.
\label{qb}
\end{equation}
Here, the complex amplitudes can be written as $\psi_0=r_0e^{i\phi_0}$ and $\psi_1=r_1e^{i\phi_1}$.
 The probabilities of measuring the qubit in the computational basis states $\ket{0}$ and $\ket{1}$ are given by
\begin{align}
\begin{array}{rc}
    P_0=\mid \psi_0 \mid ^2=r_0^2\\
    P_1=\mid \psi_1 \mid^2=r_1^2
\end{array}
  \label{qbm}
\end{align}
Applying the Hadamard gate to the state $\ket{\psi}$ transforms it as follows:
\begin{equation}
    \frac{1}{\sqrt{2}}
\begin{bmatrix}
1 & 1 \\
1 & -1
\end{bmatrix}
\begin{bmatrix}
\psi_0 \\
\psi_1
\end{bmatrix}
= \frac{1}{\sqrt{2}}
\begin{bmatrix}
\psi_0 + \psi_1 \\
\psi_0 - \psi_1
\end{bmatrix}
=
\begin{bmatrix}
\psi_0^\prime \\
\psi_1^\prime
\end{bmatrix}
\label{had}
\end{equation}
Suppose  
\[
\psi_0^\prime = r_0^\prime e^{i\phi_0^\prime}, \qquad 
\psi_1^\prime = r_1^\prime e^{i\phi_1^\prime},
\]  
then one may write  
\begin{align}
r_0^\prime e^{i\phi_0^\prime} &= a_0^\prime + i b_0^\prime, \\[6pt]
r_1^\prime e^{i\phi_1^\prime} &= a_1^\prime + i b_1^\prime,
\label{ri}
\end{align}
where  
\begin{align}
a_0^\prime &= \tfrac{1}{\sqrt{2}}\!\left(r_0 \cos \phi_0 + r_1 \cos \phi_1\right), \nonumber\\
b_0^\prime &= \tfrac{1}{\sqrt{2}}\!\left(r_0 \sin \phi_0 + r_1 \sin \phi_1\right), \nonumber\\
a_1^\prime &= \tfrac{1}{\sqrt{2}}\!\left(r_0 \cos \phi_0 - r_1 \cos \phi_1\right), \nonumber\\
b_1^\prime &= \tfrac{1}{\sqrt{2}}\!\left(r_0 \sin \phi_0 - r_1 \sin \phi_1\right).
\label{abdefs}
\end{align}

The corresponding probabilities of measuring the transformed state in the computational basis are  
\begin{align}
P_0^\prime &= {r_0^\prime}^2 
= \tfrac{1}{2}\!\left(r_0^2 + r_1^2 + 2 r_0 r_1 \cos(\phi_0 - \phi_1)\right),
\label{hadm1} \\[6pt]
P_1^\prime &= {r_1^\prime}^2 
= \tfrac{1}{2}\!\left(r_0^2 + r_1^2 - 2 r_0 r_1 \cos(\phi_0 - \phi_1)\right).
\label{hadm}
\end{align}

The phases \(\phi_0^\prime\) and \(\phi_1^\prime\) are obtained from  
\begin{align}
\phi_0^\prime &=\mathrm{atan2} \left(b_0^\prime/a_0^\prime\right), &
\phi_1^\prime &= \mathrm{atan2} \left(b_1^\prime/a_1^\prime\right).
\label{pu1}
\end{align}

\medskip



\begin{figure}
    \centering
    \includegraphics[width=0.4\linewidth]{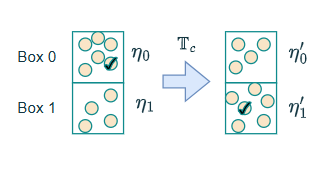}
    \caption{Schematic illustration of the system of two-box. A collection of $N$ balls is distributed between two boxes labeled $0$ and $1$, each carrying an associated phase tag $\eta_0$ and $\eta_1$. The transformation process $\mathbb{T}_c$ updates both the occupation numbers and the phase tags, mapping the initial configuration to a new one according to prescribed rules.}
    
    \label{fig:box}
\end{figure}

 

Now consider a system of two boxes (Fig.~\ref{fig:box}), labeled \(0\) and \(1\),
and a total collection of \(N\) particles, hereafter referred to as balls. The
various operations on this system are categorized into three classes:
\emph{preparation} (\(\mathbb{P}\)), \emph{transformation} (\(\mathbb{T}_c\)), and
\emph{measurement} (\(\mathbb{M}\)). Throughout the protocol, these operations
are implemented by an \emph{experimenter}, which may be either a human agent or
an automated device, and whose role is to apply the prescribed rules to the
system without introducing any additional dynamics beyond those specified by
the model.

In a typical preparation procedure $\mathbb{P}$, the experimenter begins by
selecting one ball uniformly at random from the collection of $N$ balls and
marking it for future reference. The full set of $N$ balls (including the marked
one) is then distributed between the two boxes. This allocation proceeds in one
of two ways, each chosen with equal probability:
\begin{itemize}
	\item Either $N_0 = \lfloor W_N \rfloor$ balls are randomly selected from
	the collection and placed into box $0$, with the remaining
	$N_1 = N - N_0$ balls assigned to box $1$;
	\item Or $N_0 = \lceil W_N \rceil$ balls are randomly selected from the
	collection and placed into box $0$, with the remaining
	$N_1 = N - N_0$ balls placed into box $1$.
\end{itemize}
Here, $\lfloor \cdot \rfloor$ and $\lceil \cdot \rceil$ denote the floor and ceiling functions, respectively and $W_N=\rho_0 N$. The parameter $\rho_0 \in [0,1]$ controls the expected proportion of balls
assigned to box~$0$. Throughout the selection procedure, all balls—including the
marked one—are treated as identical.

In addition, each box $k$ (with $k = 0,1$) is assigned a real-valued
\emph{phase tag} $\eta_k$. The resulting prepared state is denoted by
$\textit{B}^{N_0,N_1}_{\eta_0,\eta_1}$.

 
We now define the transformation procedure $\mathbb{T}_c$ acting on the state
$\textit{B}^{N_0,N_1}_{\eta_0,\eta_1}$. In this procedure, the experimenter first
introduces a function
\[
\tilde{N}_0 = \tilde{N}_0(N_0, N_1, \eta_0, \eta_1),
\]
and then assigns the updated occupation number $N_0'$ by choosing either
$\lfloor \tilde{N}_0 \rfloor$ or $\lceil \tilde{N}_0 \rceil$ with equal
probability.

The system is subsequently updated according to the following rule:
\begin{itemize}
	\item If $N_0' > N_0$, the experimenter randomly selects $N_0' - N_0$ balls
	from box~$1$ and transfers them to box~$0$.
	\item If $N_0' < N_0$, the experimenter randomly selects $N_0 - N_0'$ balls
	from box~$0$ and transfers them to box~$1$.
\end{itemize}

The phase tags $\eta_0$ and $\eta_1$ are also updated to new values
$\eta_0'$ and $\eta_1'$, where
\[
\eta_k' = \eta_k'(N_0, N_1, \eta_0, \eta_1), \qquad k = 0,1.
\]

Putting everything together, the transformation procedure is summarized as
\[
\mathbb{T}_c:\;
\textit{B}^{N_0,N_1}_{\eta_0,\eta_1}
\;\longrightarrow\;
\textit{B}^{N_0',N_1'}_{\eta_0',\eta_1'} .
\]

The \emph{measurement} procedure, denoted by $\mathbb{M}$, is implemented as
follows. In this
process, the experimenter places all balls into the box that contains the
marked ball, and the \emph{measurement outcome} is identified with the label of
that box. Both phase tags are then reset to zero.

Accordingly, the post-measurement state becomes
\begin{itemize}
	\item $\textit{B}^{N,0}_{0,0}$ if the marked ball is in box~$0$;
	\item $\textit{B}^{0,N}_{0,0}$ if the marked ball is in box~$1$.
\end{itemize}
Suppose the experimenter prepares an initial state
$\textit{B}^{N_0,N_1}_{\phi_0,\phi_1}$, with the identifications
$\eta_0 = \phi_0$ and $\eta_1 = \phi_1$. Here $\phi_0$ and $\phi_1$ are the phases
of the complex probability amplitudes $\psi_0$ and $\psi_1$ defined in
Eq.~\eqref{qb}. The occupation numbers $N_0$ and $N_1$ are obtained through the
preparation procedure $\mathbb{P}$ by choosing
$\rho_0 = r_0^2$, so that $W_N= r_0^2 N$, where $r_0^2$ is the probability
of obtaining the outcome $\ket{0}$ from the quantum state in
Eq.~\eqref{qb}.

If the experimenter then applies the measurement procedure $\mathbb{M}$, the
probability of obtaining outcome $0$ in the limit of large $N$ is
\begin{equation}
	\lim_{N\to\infty}
	\frac{\langle N_0 \rangle}{N}
	= \lim_{N\to\infty}
	\frac{\lfloor {N}_0 \rfloor + \lceil {N}_0 \rceil}{2N}
	= r_0^2,
	\label{hm10}
\end{equation}
while the probability of obtaining outcome $1$ becomes
\begin{equation}
	\lim_{N\to\infty}
	\left( 1 - \frac{\langle N_0 \rangle}{N} \right)
	= r_1^2.
	\label{hm20}
\end{equation}


Now suppose the experimenter starts from the same initial state
$\textit{B}^{N_0,N_1}_{\phi_0,\phi_1}$ and applies a specific transformation to this state.
This transformation is designed to reproduce the action of the Hadamard coin
(Eqs.~\eqref{hadm1}, \eqref{hadm}, and \eqref{pu1}) within the present classical
box and balls framework.

In this $\mathbb{T}_c$ procedure, the experimenter follows the transformation
scheme defined below. First, the quantity
\begin{equation}
	\tilde{N}_0
	= \frac{N_0 + N_1}{2}
	+ \sqrt{N_0 N_1}\,\cos(\phi_1 - \phi_0)
	\label{ntilde}
\end{equation}
is introduced, together with the auxiliary functions
\begin{align}
	B_k &= \sqrt{\frac{N_0}{N}}\,\sin \phi_0
	+ (-1)^k \sqrt{\frac{N_1}{N}}\,\sin \phi_1, \\
	A_k &= \sqrt{\frac{N_0}{N}}\,\cos \phi_0
	+ (-1)^k \sqrt{\frac{N_1}{N}}\,\cos \phi_1.
\end{align}

The updated phase tags associated with the boxes are then determined by
\begin{equation}
	\eta_k'(N_0,N,\phi_0,\phi_1)
	= \mathrm{atan2}\left(\frac{B_k}{A_k}\right),
	\label{etap}
\end{equation}
and the experimenter sets the updated number of balls in box~$0$, denoted
$N_0'$, to either
\[
\left\lfloor \frac{N}{2} + \sqrt{N_0 N_1}\cos(\phi_1 - \phi_0) \right\rfloor
\quad \text{or} \quad
\left\lceil \frac{N}{2} + \sqrt{N_0 N_1}\cos(\phi_1 - \phi_0) \right\rceil,
\]
each choice occurring with equal probability. The method by which the experimenter obtained this updated state has been described by the general method of the transformation process $\mathbb{T}_c$ introduced previously.

The specific forms of $\tilde{N}_0$, $A_k$, and $B_k$ in the transformation
$\mathbb{T}_H$ are chosen such that, in the limit $N \to \infty$, the probability
of obtaining the measurement outcome $0$ satisfies
\begin{equation}
	\lim_{N\to\infty}
	\frac{\langle N_0^\prime \rangle}{N}
	= \lim_{N\to\infty}
	\frac{\lfloor \tilde{N}_0 \rfloor + \lceil \tilde{N}_0 \rceil}{2N}
	= {r_0^\prime}^2,
	\label{hm1}
\end{equation}
while the probability of outcome $1$ becomes
\begin{equation}
	\lim_{N\to\infty}
	\left( 1 - \frac{\langle N_0^\prime \rangle}{N} \right)
	= {r_1^\prime}^2.
	\label{hm2}
\end{equation}
In addition, the phase-tag updates are also satisfy
\begin{equation}
	\lim_{N\to\infty}
	\langle \eta_k'(N_0,N,\phi_0,\phi_1) \rangle
	= {\phi_k}'.
\end{equation}

As $N \to \infty$, the measurement-outcome probabilities generated by this
classical procedure converge to those produced by the Hadamard transformation on the quantum state of Eq. \eqref{qb} .

For finite $N$, deviations are unavoidable; however, by taking $N$ sufficiently
large, such quantum operations can be approximated with arbitrary precision
within the present classical framework.


Moreover, this method can be generalized to simulate other single-qubit quantum coin operations, as expressed in Eq.~\eqref{general_coin}. One can obtain the generalized forms of \(\tilde{N}_0\) and \(\eta_i'\) in the corresponding transformation process as follows :
 \begin{align}
\begin{array}{rc}
   \tilde{N}_0 =N_0\cos^2\theta+N_1\sin^2\theta&\\
  &\hspace{-5.6em}+2\sqrt{N_0N_1}\sin\theta\cos\theta\cos(\phi_0-\phi_1+ \xi-\zeta),  \\
  \end{array}
  \label{gc_nt}
  \end{align}
 
 \begin{align}
\eta_k^\prime=\mathrm{atan2} \left(\frac{\mathbb{B}_k}{\mathbb{A}_k}\right),
\label{gc_pt}
 \end{align}
 where,
$$\mathbb{B}_k=\sqrt{\frac{N_0}{N}}\cos\theta\sin (\xi+\phi_0)+(-1)^k\sqrt{\frac{N_1}{N}}\sin\theta\sin(\zeta+ \phi_1),$$
$$\mathbb{A}_k=\sqrt{\frac{N_0}{N}}\cos\theta\cos(\xi+ \phi_0)+(-1)^k\sqrt{\frac{N_1}{N}}\sin\theta\cos(\zeta+ \phi_1).$$


 Finally, we introduce the boxes-and-balls model corresponding to a 1D discrete-time quantum walk (DTQW) with a generalized $SU(2)$ coin [Eq.~\eqref{general_coin}], and an initial state of the form:
 \begin{figure}
 	\centering
 	\includegraphics[width=1.0\linewidth]{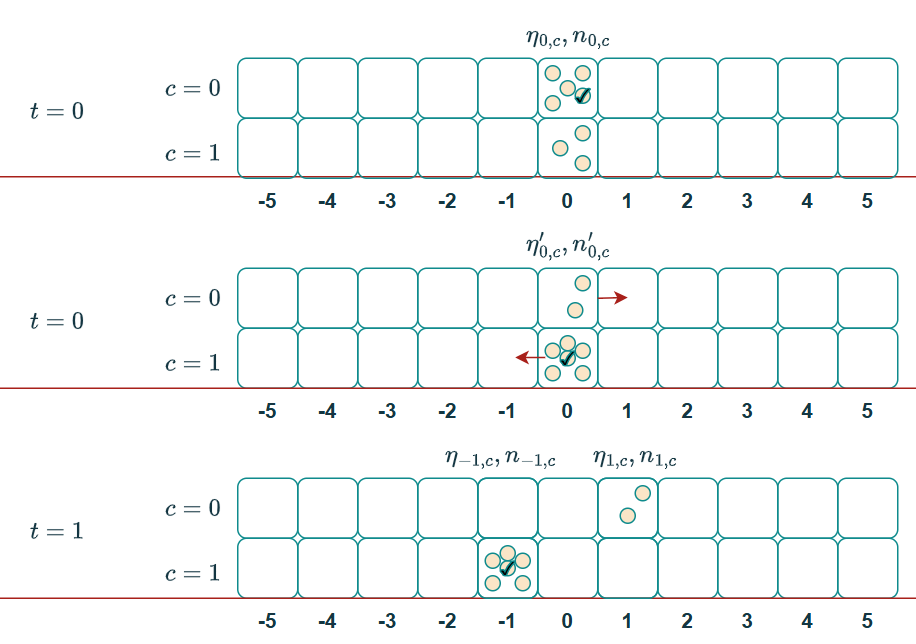}
 	\caption{Schematic illustration of the  boxes-and-balls model for a discrete-time quantum walk. Each lattice site $x\in\mathbb{Z}$ contains two boxes labeled by $c=0,1$, with associated phase tags $\eta_{x,c}$. At time $t=0$, the $N$ balls (one of which is marked) are distributed according to the preparation process $\mathbb{P}$. The transformation $\mathbb{T}_c$ updates both the occupation numbers $N_{0,c}$ and the phase tags, followed by the conditional shift that moves the balls and associated phase tags to left or right depending on $c$. The configuration at $t=1$ illustrates the resulting redistribution after one full step. }
 	\label{qwf}
 \end{figure}
\begin{figure}
    \centering
    \includegraphics[width=0.9\linewidth]{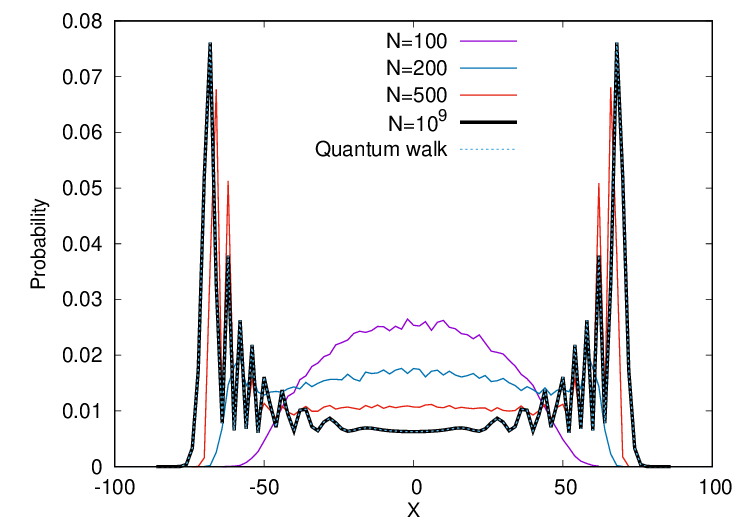}
    \caption{Probability distributions arising from the $\mathbb{M}$ process in the  boxes-and-balls model of the Hadamard coined quantum walk, with initial state $\frac{1}{\sqrt{2}}\ket{0} - \frac{i}{\sqrt{2}}\ket{1}$ and at time $t = 100$, are shown for different total numbers of particles $N$. 
     As $N$ increases, the distribution approaches that of the ideal Hadamard quantum walk, indicated by the dotted line. }
    \label{fn}
\end{figure}

\begin{equation}
 \ket{\psi(x,t=0)}=\left(\psi_0\ket{0}+\psi_1\ket{1}\right)\otimes\ket{x=0}  
 \label{qwi}
\end{equation}

In this setting, the dynamics is implemented through the three processes $\mathbb{P}$, $\mathbb{T}_c$, and $\mathbb{M}$ within a generalized framework. To each lattice site $x \in \mathbb{Z}$ we assign two boxes and a phase tag to each box, labeled $(x,c)$ and $\eta_{x,c}$ respectively, with $c\in\{0,1\}$ (Fig.~\ref{qwf}). We begin with a collection of $N$ balls, one of which is randomly selected and marked by the experimenter. The full set of $N$ balls is then distributed among the boxes according to a preparation protocol that will be specified later.

At any time $t$, the state of the system is completely characterized by the nonnegative integers
\[
N_{x,c}(t)=\text{number of balls in box }(x,c), 
\qquad c\in\{0,1\},
\]
together with a phase tag $\eta_{x,c}(t)\in[0,2\pi)$ associated with each box $(x,c)$. The occupation numbers $N_{x,c}$ satisfy the global constraint
\begin{equation}
	\sum_{x\in\mathbb{Z}}\sum_{c=0}^1 N_{x,c}(t)=N,
	\qquad \forall\,t.
\end{equation}

At time \(t=0\), only the two boxes at the origin contains the balls ; all others are empty:
\[
N_{x,c}(0)=0
\quad\text{for}\quad
(x,c)\neq(0,0),(0,1).
\]
Within the origin boxes, we set
\[
N_{0,0}(0)=
\begin{cases}
\lfloor r_0^2N\rfloor, & \text{with probability }1/2,\\[6pt]
\lceil r_0^2N\rceil,  & \text{with probability }1/2,
\end{cases}
\]
\[
N_{0,1}(0)=N - N_{0,0}(0),
\]
and assign phases $\eta_{0,0}(0)=\phi_0$, $\eta_{0,1}(0)=\phi_1$ so as to encode the initial coin–position state of Eq.\ref{qwi}. All other phase tags are set to zero, i.e.,
\[\eta_{x,c}=0
\quad\text{for}\quad
(x,c)\neq(0,0),(0,1).\]
Now for each discrete time step $t\to t+1$, we perform two sequential sub‐steps:
\begin{figure}
    \centering
    \includegraphics[width=0.7\linewidth]{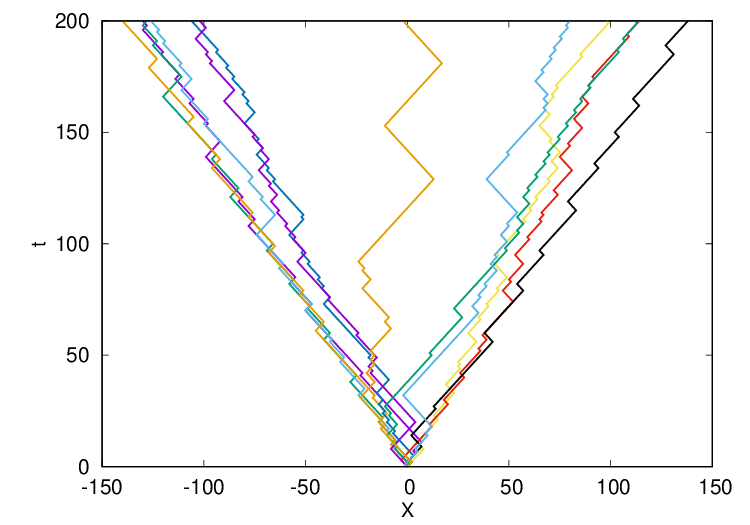}
    \caption{Snapshot of the trajectories of the marked ball in the  boxes-and-balls model ($N=10^9$) of the 1D Hadamard quantum walk, initialized in the state $\frac{1}{\sqrt{2}}\ket{0}-\frac{i}{\sqrt{2}}\ket{1}$, shown at time $t=200$}
    \label{snapshot}
\end{figure}
  \paragraph*{Coin–Transformation \(\mathbb{T}_c\):} 
    At every site $x$ with $N_{x,0}(t)+N_{x,1}(t)>0$, we apply a $\mathbb{T}_c$ process on the two boxes labeled by $(x,0)$ and $(x,1)$ with $\tilde{N}_{x,0}$, $\eta_{x,k}$ and $\eta_{x,k}^\prime$ will play the role of $\tilde{N}_{0}$, $\phi_k$ and  $\eta_{k}^\prime$ respectively as given in Eq. \eqref{gc_nt} and Eq. \eqref{gc_pt}. This is a stochastic mixing rule of the two boxes at each site that mirrors the quantum “coin” operation. Concretely, box‐occupancies and phases are updated to new values $N'_{x,c}(t)$ and $\eta'_{x,c}(t)$ so as to reproduce, in the large–$N$ limit, the SU(2) coin operation of Eq.~\eqref{general_coin}.
  \paragraph*{Conditional Shift \(\mathbb{S}\):}
    After the \(\mathbb{T}_c\), all balls in box $c=0$ at site $x$ move to box $0$ of site $x+1$, while those in $c=1$ move to box $1$ of site $x-1$. Phases are also carried along unchanged:
    \begin{align}
      N_{x+1,0}(t+1)&=N'_{x,0}(t), & \eta_{x+1,0}(t+1)&=\eta'_{x,0}(t),\\
      N_{x-1,1}(t+1)&=N'_{x,1}(t), & \eta_{x-1,1}(t+1)&=\eta'_{x,1}(t),
    \end{align}
    and all other box‐occupancies reset to zero.

The \( \mathbb{M} \) process is then implemented within this generalized framework as follows.

\paragraph*{Measurement (\(\mathbb{M}\) process):}
After $t$ steps of iterating \(\mathbb{T}_c\) and \(\mathbb{S}\), we perform the measurement process $\mathbb{M}$: we identify the box $(x_{\rm m},c_{\rm m})$ containing the marked ball, move all balls into that box, and record the site index $x_{\rm m}$. The resulting histogram of recorded $x_{\rm m}$ over many trials reproduces the probability distribution of the quantum walk at time $t$ in the large $N$ limit [Fig. \ref{fn}].\\


The dynamics can be viewed from a different perspective also: by iterating \(\mathbb{T}_c\) and \(\mathbb{S}\) for $t$ steps, the marked particle undergoes a classical random walk [Fig. \ref{snapshot}] whose large–$N$ distribution converges to that of the quantum walker’s position distribution $|\psi(x,t)|^2$.

In the box–ball formulation, essentially the boxes interact through the exchange of balls. This idea of ball relocation can be extended to the level of individual particle dynamics by employing an active spin model.

Active spin models \cite{Solon2013PRL,Solon2015PRE,Bandyopadhyay2024PRE,Karmakar2023PRE,Haldar2023PRE} describe collections of self-propelled particles carrying an internal two-state degree of freedom (spin), whose dynamics combines stochastic spin flips with biased motion on a lattice. Such models are known to exhibit rich nonequilibrium phenomena, including collective motion, phase separation, and long-range order in low dimensions.  The primary motivation for such models is twofold: first, to
identify the minimal ingredients required for collective motion; and second,
to understand how nonequilibrium driving circumvents equilibrium constraints
on ordering and phase transitions.
Building on the box--ball formulation of the discrete-time quantum walk,
we introduce an active spin model obtained directly from this framework.\\

In this construction, each ball is reinterpreted as an active particle and is
endowed with an Ising spin variable $s=\pm 1$ and a real-valued phase
$\eta_x^{\,s}$, inherited from the box index $c\in\{0,1\}$ and the associated
phase tag $\eta_{x,c}$ of the box--ball representation. The dynamics consists of two coupled processes: (i) \emph{spin-flip and phase update} events that modify the internal state as well as the phase of a particle, and (ii) \emph{active hopping} on the lattice, in which the direction and rate of motion depend on the spin. The spin-flip and hopping rates are chosen so that the particle density produces the probability distribution of the discrete time quantum walker when the number of particles are sufficiently large.

The explicit definition of the spin variables, flip rates, and hopping rules is introduced below. For the moment, the essential point is that the classical active dynamics is engineered to mirror the unitary coin–shift structure of the quantum walk.\\

We consider $N$ particles carrying Ising spins $s=\pm 1$ on a one-dimensional
lattice of $L$ sites. In addition to the spin degree of freedom, each particle
carries a real-valued phase $\eta_x^{\,s}$ that depends only on its lattice
position $x$ and spin $s$. Each particle hops to its right and left neighboring
sites at rates $D(1+\varepsilon s)$ and $D(1-\varepsilon s)$, respectively.
There is no exclusion between particles, and we denote by
$n_x^{\pm1}$ the number of $\pm$ spins on site $x$, so that the local density
is given by $\rho_x = n_x^{+1} + n_x^{-1}$.
The particles also flip their spins: on site $x$ each spin flips at rate
$$R_{x,s}=\frac{s}{2}(1+\frac{s\delta_x}{\lvert \delta_x \rvert})\frac{\delta_x}{n_x^s},$$
Here $\delta_x =  n_x^{+1}-\tilde{n}_x^{+1} $, where $\tilde{n}_x^{+1}$ is defined
by the same expression as $\tilde{N}_0$ in Eq.~\eqref{gc_nt}, with the local
variables $n_x^{+1}$, $n_x^{-1}$, $\eta_x^{+1}$, and $\eta_x^{-1}$ playing the
roles of $N_0$, $N_1$, $\phi_0$, and $\phi_1$, respectively.
\begin{align}
	\begin{array}{rc}
	\tilde{n}_x^{+1}=n_x^{+1}\cos^2\theta+n_x^{-1}\sin^2\theta &\\
	&\hspace{-9.6em}+2\sqrt{n_x^{+1}n_x^{-1}}\sin\theta\cos\theta\cos(\eta_x^{+1}-\eta_x^{-1}+ \xi-\zeta),  \\
	\end{array}
	\label{gc_nt}
\end{align}

 The updated phases ${\eta_x^{+1}}'$ and ${\eta_x^{-1}}'$ correspond to $\eta_{x,0}'$ and $\eta_{x,1}'$
in the box-and-balls construction.
\begin{align}
	{\eta_x^{s}}'=\mathrm{atan2} \left(\frac{\mathbb{B}_s}{\mathbb{A}_s}\right),
	\label{gc_pt}
\end{align}
where,
$$\mathbb{B}_s=\sqrt{\frac{n_x^{+1}}{N}}\cos\theta\sin (\xi+\eta_x^{+1})+s\sqrt{\frac{n_x^{-1}}{N}}\sin\theta\sin(\zeta+ \eta_x^{-1}),$$
$$\mathbb{A}_s=\sqrt{\frac{n_x^{+1}}{N}}\cos\theta\cos(\xi+ \eta_x^{+1})+s\sqrt{\frac{n_x^{-1}}{N}}\sin\theta\cos(\zeta+ \eta_x^{-1}).$$

For $D=0.5$ and $\varepsilon=1$, i.e., when the hopping is completely biased, if the initial particle configuration matches the corresponding initial configuration of the box-and-ball model, the resulting density distribution reproduces the probability distribution of the discrete-time quantum walk in the large-$N$ limit. This behavior has been confirmed by direct numerical simulations.

Recently, it has been shown in the literature that suitably designed active matter systems can exhibit emergent features characteristic of quantum dynamics, including interference-like effects \cite{Maes2022}, the ability to tunnel through potential barriers \cite{TeVrugt2023}, Bose--Einstein condensation~\cite{Meng2021}, Schr\"odinger-type dynamics in polar liquids~\cite{Souslov2017},  spin--orbit coupling~\cite{Loewe2018ac}, topological effects~\cite{Palacios2021NatComm} and  Hall viscosities~\cite{Han2021,Markovich2021} . Theses allows for the simulation of quantum phenomena in classical, controlled environments.
Motivated by these developments, the present construction aims to provide a minimal, lattice-based active model whose collective dynamics reproduces the behavior of a discrete-time quantum walk. Since quantum walks are known to be universal simulators of quantum dynamics (any quantum circuit can be encoded in an appropriate
quantum-walk architecture), this mapping highlights the expressive power of active matter as a classical platform for simulating quantum-like phenomena.




In this letter, we develop classical interacting particle models that simulate one-dimensional discrete-time quantum walks with a generalized coin. The framework can be naturally extended to higher-dimensional lattices and complex networks by associating multiple boxes (directions) with each node and introducing appropriate transformation and shift operations.

The strength of the present model lies in its simplicity and the valuable insights it offers. 
Successful extensions of this approach could enable the design of DTQW-based algorithms through purely particle-base reasoning. Various active spin models can themselves simulate a wide variety of collective dynamics. As this generalised active spin model can reflect the behavior of quantum walk, they can lead to the development of novel quantum algorithms for simulating various collective phenomena in active systems, as well as quantum-inspired classical strategies. in this regard it is important to notice that the present work demonstrates that Discrete-time quantum walks are themselves a quantum algorithm for simulating a class of classical active spin dynamics in the limit where the number of particles tends to infinity.

Quantum walks have found successful applications across a wide range of domains \cite{qs,cd,ph1,ph2,ph3,mof,mof2,cmc2,cmc3,tkit,kh,nao,yam}, from quantum search algorithms \cite{qs} to simulating relativistic quantum dynamics \cite{cmc2,cmc3} and modeling quantum active particles \cite{yam}. 

The present study establishes a connection between the generalised active spin model and all these domains where quantum walks are effectively applied.

In contrast to classical random walks, quantum walks do not permit a complete record of the walker’s trajectory due to the measurement-induced collapse of the quantum state, which disrupts its coherent evolution. As a result, traditional notions like first-passage time (FPT) are not directly applicable, and the first-detected-passage time (FDPT) becomes the relevant quantity of interest \cite{Shukla2025,frid}. However, within the box model framework of DTQWs, it becomes possible to define both FPT and FDPT meaningfully. This dual capability offers a novel platform for examining first passage times in quantum system.

Conceptually, our approach belongs to the class of classical frameworks that reproduce quantum behavior without assigning a dynamical role to the wave function \cite{holl,poir,parl,shiff, sebens,wiseman,schol}, providing a simple and accessible route toward deeper structural insights into quantum theory.


\textit{Acknowledgments:} The author thanks Nicolas Gisin, Parongama Sen, and Arghya Das for valuable discussions.

\end{document}